# INTEGRATING HYPERTENSION PHENOTYPE AND GENOTYPE WITH HYBRID NON-NEGATIVE MATRIX FACTORIZATION

**Yuan Luo**  Yuan.luo@Northwestern.edu
Department of Preventive Medicine, Feinberg School of Medicine
Northwestern University
Chicago, IL, USA

**Chengsheng Mao**  chengsheng.mao@Northwestern.edu
Department of Preventive Medicine, Feinberg School of Medicine
Northwestern University
Chicago, IL, USA

**Yiben Yang**  yiben.yang@Northwestern.edu
Department of Preventive Medicine, Feinberg School of Medicine
Northwestern University
Chicago, IL, USA

**Fei Wang**  feiwang03@gmail.com
Department of Healthcare Policy & Research, Weill Cornell Medicine
Cornell University
New York, NY, USA

**Faraz S. Ahmad**  faraz.ahmad@Northwestern.edu
Department of Preventive Medicine, Feinberg School of Medicine
Northwestern University
Chicago, IL, USA

**Donna Arnett**  donna.arnett@uky.edu
College of Public Health
University of Kentucky
Lexington, KY, USA

**Marguerite R. Irvin**  irvinr@uab.edu
Department of Epidemiology
University of Alabama at Birmingham
Birmingham, AL, USA

**Sanjiv J. Shah**  sanjiv.shah@Northwestern.edu
Department of Medicine, Feinberg School of Medicine
Northwestern University
Chicago, IL, USA

## Abstract

Hypertension is a heterogeneous syndrome in need of improved subtyping using phenotypic and genetic measurements so that patients in different subtypes share similar pathophysiologic mechanisms and respond more uniformly to targeted treatments. Existing machine learning approaches often face challenges in integrating phenotype and genotype information and presenting to clinicians an interpretable model. We aim to provide informed patient stratification by introducing Hybrid Non-negative Matrix Factorization (HNMF) on phenotype and genotype matrices. HNMF simultaneously approximates the phenotypic and genetic matrices using different appropriate loss functions, and generates patient subtypes, phenotypic groups and genetic groups. Unlike previous methods, HNMF approximates phenotypic matrix under Frobenius loss, and genetic matrix under Kullback-Leibler (KL) loss. We propose an alternating projected gradient method to solve the approximation problem. Simulation shows HNMF converges fast and accurately to the true factor matrices. On real-world clinical dataset, we used the patient factor matrix as features to predict main cardiac mechanistic outcomes. We compared HNMF with six different models using phenotype or genotype features alone, with or without NMF, or using joint NMF with only one type of loss. HNMF significantly outperforms all comparison models. HNMF also reveals intuitive phenotype-genotype interactions that characterize cardiac abnormalities.

## 1. Introduction

Precision medicine aims to utilize information from multiple modalities—including phenotypic and genetic measurements—to develop an individualized and comprehensive view of a patient's pathophysiologic progression, to identify unique disease subtypes, and to administer personalized therapies (Kohane, 2015). Existing efforts are often based on only a selected set of biomarkers. The rapid growth of phenotypic and genetic data for many common diseases poses technical challenges for subtyping them, due to the high dimensionality of data, diversity of data types, uncertainty and missing data. However, the rapid growth of multiple data modalities, when linked to the right patients, may provide a prismatic view of the underlying pathophysiology of these diseases and offers a basis for meaningful subtyping of these patients.

Hypertension is an example of a complex, heterogeneous clinical syndrome characterized by elevated blood pressure. Although typically considered a single disease, primary hypertension (i.e., essential hypertension) is in fact a heterogeneous group of subtypes with varying etiologies and pathophysiology. This common form of hypertension is highly prevalent and is polygenic in nature. However, genetic studies of hypertension have focused primarily on analyzing single variants at a time and then ranking them in terms of significance, as has been done in several genome-wide association studies (see (Poulter, Prabhakaran, & Caulfield, 2015) for a review). However, it is more likely that genetic variants interact with each other to increase susceptibility to disease. Furthermore, genetic variants interact with phenotypic risk factors to further promote the development of diseases such as hypertension. With the growing availability of high throughput genotyping and phenotyping data (such as through NIH/NHLBI TOPMed program), the need for integrating both data modalities is becoming increasingly pressing. Thus, it is critical to develop a methodology to combine phenotypic and genetic data when clustering patients for the identification of novel subtypes of disease. Such work could help identify novel molecular and pathophysiological pathways of disease and also may identify subgroups of patients who are more homogeneous in their response to specific therapies.

Major contributions of this paper are: 1) Aiming to provide informed patient stratification, we propose Hybrid Non-negative Matrix Factorization (HNMF) that approximates phenotype and genotype matrices using different appropriate loss functions, instead of single loss function in previous joint NMF methods. 2) We use simulation to show HNMF converges fast and accurately to true factor matrices, and we use a real-world clinical dataset to show HNMF-generated patient factor matrix is more effective in predicting indices of cardiac mechanics compared to multiple non-NMF, NMF and joint NMF based methods. 3) We show that HNMF-generated group matrices lead to insights on phenotype-genotype interactions that characterize cardiac abnormalities.

From the clinical perspective, there have been only a few previous studies that have examined the clustering of hypertensive patients. Katz et al. applied model-based clustering to a cohort of 1,273 hypertensive individuals, using only phenotypic data as features (Katz et al., 2017). Study participants were clustered into 2 distinct groups that differed markedly in clinical characteristics, cardiac structure/function, and indices of cardiac mechanics. Guo et al. (Guo et al., 2017) used K-means clustering of phenotypic data (clinical and blood pressure characteristics) and found 4 groups of interest. However, neither of these studies utilized genetic data, which could have provided an additional important dimension to the clustering of hypertension, particularly when combined with phenotypic data.

From the method perspective, Non-negative Matrix Factorization (NMF) refers to the set of problems on approximating a non-negative matrix as the product of several non-negative matrices. The problem has become popular since Lee and Seung's Nature paper (D. D. Lee & Seung, 1999), where they form a nonnegative matrix by concatenating the set of pixel intensity vectors stretched from human facial images. After factorizing such matrix into the product of two matrices, they found that one matrix can be interpreted as the set of image basis with part based representation of human faces, and the other matrix is the coefficients if we were to reconstruct the face image from those bases. Because of the non-negativity constraints, NMF is not a convex problem and they developed a multiplicative update algorithm to obtain a stationary solution, with provable convergence of the algorithm (Daniel D Lee & Seung, 2001).

Since then researchers have been working on NMF from various aspects. Ding and Simon (C. Ding, He, & Simon, 2005) showed that there is some equivalence between NMF and Kmeans/spectral clustering and claim NMF can be used for data clustering. Ding et al. (C. Ding, Li, Peng, & Park, 2006) further developed a t-NMF approach that can perform co-clustering on both matrix columns and rows. They also discussed the various NMF variants (C. H. Ding, Li, & Jordan, 2010). Dhillon et al. (Sra & Dhillon, 2006) extended NMF to the case when the matrix approximation loss is measured by Bregman divergence, which is a much more general loss with both Frobenius norm and KL divergence (which are discussed in (Daniel D Lee & Seung, 2001)) as its special cases. On the solution procedure aspect, multiplicative updates have been recognized for its slow convergence and poor quality. Lin (Lin, 2007) proposed a projected gradient approach for NMF. Kim and Park (J. Kim & Park, 2011) also proposed an active set type of method called principal block pivoting to solve the NMF problem.

NMF is a highly effective unsupervised method to cluster similar patients (Hofree, Shen, Carter, Gross, & Ideker, 2013; Yuan Luo, Xin, Joshi, Celi, & Szolovits, 2016) and sample cell lines (Müller et al., 2008), and to identify subtypes of diseases (Collisson et al., 2011). Conventional NMF can only model either phenotypic measurements (e.g. using Frobenius loss) or genetic variants (e.g., using KL loss) but not both. Recent studies have investigated methods for joint matrix factorization, serving the purpose of meta-analysis (Wang, Zheng, & Zhao, 2014), multi-view clustering (Liu, Wang, Gao, & Han, 2013) or imposing multiple characterization of documents (H. Kim, Choo, Kim, Reddy, & Park, 2015). However, these methods focus on using Frobenius loss to measure approximation accuracy of multiple matrices, and cannot readily integrate phenotypic measurements and genetic variants where approximating the two matrices admit different types of loss functions. Gunasekar et al. (Gunasekar et al., 2016) proposed collective matrix factorization based on the Bregman divergence framework to integrate multi-source EHR phenotyping data, implemented KL-divergence as matrix approximation loss and experimented on discrete diagnosis and medications data.

In theory, both KL divergence and Frobenius loss are special cases of Bregman divergence, but care needs to be taken when materializing the theoretical framework to the concrete case of hybrid genotypic and continuous phenotypic data. Challenges include how to derive useful genetic variant information from terabytes of whole exome sequencing data, how to filter deleterious variants, how to properly implement HNMF with missing continuous data, etc. Our paper is one such concrete materialization to integrate phenotypic and genotypic information for patient subtyping. Addressing both the clinical and methodological challenges, we propose the model of Hybrid Non-negative Matrix Factorization (HNMF) which models the approximations of phenotypic and genetic matrices under Frobenius loss and KL loss respectively. We develop an alternating project gradient descent method for optimizing the HNMF objective, and demonstrate its fast convergence and effectiveness in integrating both the phenotypic and genetic data using both simulated and real-world studies.

## 2. Materials and Methods

We develop a hybrid matrix factorization method to integrate both phenotypic and genetic features. The model applies non-negative matrix factorization to discover groups of phenotypic variables and genetic variants simultaneously that collectively and interactively characterize the groups of the patients. The approximation error is measured using Frobenius loss for the phenotypic matrix, and KL loss for the genetic matrix; hence we name our algorithm the Hybrid Non-negative Matrix Factorization (HNMF).

### 2.1 Workflow of the study

We first outline the workflow of the study in Figure 1. This study considers two types of patient data: phenotypic measurements and genetic variants.

We first impute missing values in the phenotypic variables. For genetic variants, we first annotate the variants and then keep those variants that are likely gene disruptive (LGD). The pre-processed phenotypic measurements and genetic variants are then used as input to our HNMF algorithm. The patient factor matrix is then used as the feature matrix to perform regression analysis to predict main cardiac mechanistic outcomes. We next explain each step in detail.

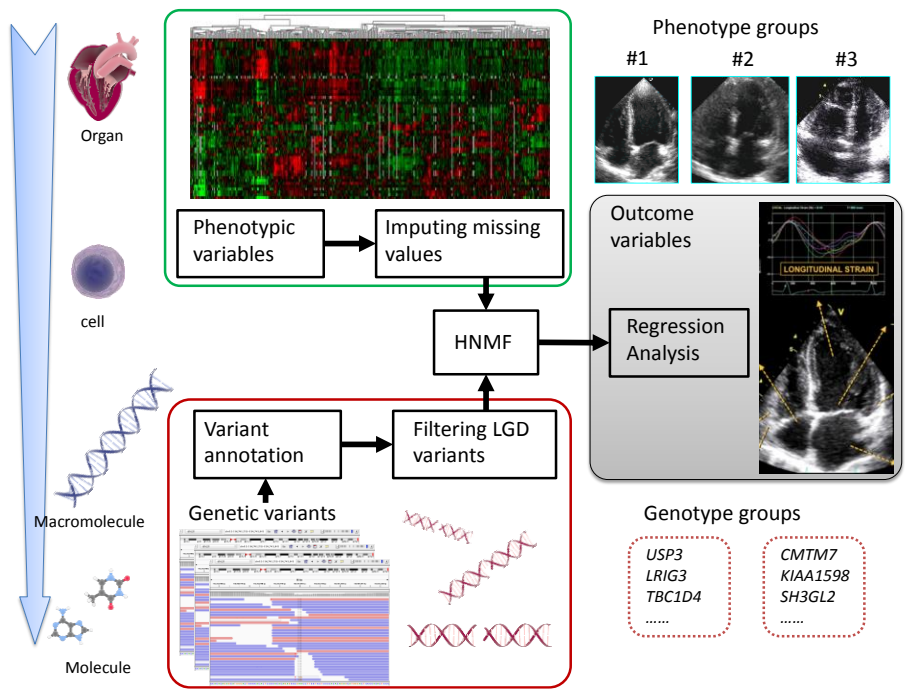

Figure 1: Study workflow. HNMF stands for hybrid non-negative matrix factorization. LGD stands for likely gene disruptive.

## 2.2 Cohort construction and data collection

Our cohort comes from the Hypertension Genetic Epidemiology Network (HyperGEN) study. HyperGEN, part of the NIH Family Blood Pressure Program, is a cross-sectional study consisting of individuals with hypertension, their siblings and offspring, and a random sample of normotensives, all recruited from 4 cities in the United States (Williams et al., 2000). We focus on the African American participants (660 total), for whom we have both the phenotypic data (e.g., vitals) and whole exome sequencing (WES) data. We used two measurements from the echocardiograms that are main reflectors of systolic (longitudinal strain) and diastolic (septal e' velocity) cardiac mechanics as outcome variables (Table 1) (Mitter, Shah, & Thomas, 2017; Mor-Avi et al., 2011). As opposed to conventional cardiac function measures such as ejection fraction, indices of cardiac mechanics obtained by speckle-tracking echocardiography are more sensitive measures of intrinsic cardiomyocyte function (Shah et al., 2014). Furthermore, indices of cardiac mechanics are thought to be subclinical measures of myocardial dysfunction that occur during the transition from risk factors (e.g., hypertension, obesity, diabetes, renal disease) to overt heart failure (Selvaraj et al., 2016). WES identifies the variants found in the coding region of genes (exons). In order to accurately and consistently call variants from across all datasets, we adopt the GATK framework (DePristo et al., 2011) for a standardized processing of WES data.

Table 1: Outcome variables reflecting cardiac mechanics

| Outcome | Description |
| --- | --- |
| Septal e' velocity | Left ventricular early diastolic relaxation velocity, measured at the septal mitral annulus in the apical 4-chamber view. Lower values reflect slower left ventricular relaxation and worse diastolic function. |
| Longitudinal strain | Left ventricular longitudinal strain measured in the apical 4-chamber view, a marker of subendocardial longitudinal systolic function. Lower absolute values reflect worse systolic function. |

## 2.3 Imputation on phenotypic variables

Biomedical, epidemiological and clinical data often contain missing values for test results, some due to issues during data acquisition and archiving, but others due to the fact that clinicians do not order certain tests based on patient-specific diagnostic and treatment course. The missing percentage of the phenotypic variables considered in our study ranges from 0% to 37%. We had our cardiologist colleagues pick rather inclusively 129 phenotypic variables that can characterize the hypertension risk and cardiac physiology of the patients. We are rather tolerant on missing rate in order to retain as many variables as possible. Nevertheless, only 13/129 (10%) of the phenotypic variables had missingness > 10%. Six of these variables with missingness > 10% (including those with missingness > 30%) were phenotypes related to mitral inflow, which characterize diastolic function. Given the importance of diastolic dysfunction (i.e., abnormal cardiac relaxation and/or reduced cardiac compliance) in the setting of hypertension, we chose to retain these variables because of their clinical importance. We use the Multivariate Imputation by Chained Equations (MICE) algorithm to perform the imputation. This approach assumes a conditional model for each variable to be imputed, with the other variables as possible predictors (van Buuren & Groothuis-Oudshoorn, 2011). The term chained equation comes from the adoption of a Gibbs sampler, which is an iterative Markov Chain Monte Carlo (MCMC) algorithm. Previous studies (e.g., (Y. Luo, Szolovits, Dighe, & Baron, 2016)) showed that even at the presence of high missing rate (over 50%), MICE imputation may still render clinically useful information to predict patient outcome due to redundant information in phenotypic variables.

## 2.4 Annotation-based variant filtration and LGD variant detection

We next used the ANNOVAR toolkit (K. Wang, Li, & Hakonarson, 2010) to comprehensively annotate called variants with a wide array of information, including their hosting gene (using several gene models such as RefSeq, UCSC Known Gene, Gencode (Harrow et al., 2012)); the variant function; its predicted pathogenicity according to PolyPhen2 (Adzhubei, Jordan, & Sunyaev, 2013), SIFT (Kumar, Henikoff, & Ng, 2009), CADD (Kircher et al., 2014), and other meta predictors; its minor allele frequency among the 1000 Genomes populations and ExAC (Lek et al., 2016); and its phenotype associations according to ClinVar, and HGMD (Stenson et al., 2012).

To address issues of reference mis-annotation, we resort to the recent-ly released Exome Aggregation Consortium (ExAC) exome dataset (Lek et al., 2016), which aims to aggregate exome sequencing data from a wide range of large-scale sequencing projects including the cohorts of Myocardial Infarction Genetics Consortium, Swedish Schizophrenia & Bipolar Studies and The Cancer Genome Atlas (TCGA). We filter out those variants whose allele frequencies are observed to be over 90% among the 60,706 individuals aggregated by ExAC. We also apply a similar 90% filtering threshold on the alternate allele frequency in our cohort. We further focus on likely gene disruptive (LGD) variants, which include frame-shift insertion, frame-shift deletions, nonsense variants, and splice site alterations. We have 6430 gene features for our cohort, 660 subjects. We follow the common practice and exclude the genes that have very rare variants (<10 subjects) or very frequent variants (> 50% of the subjects), resulting in 1481 genes. We then follow the com-mon approach of gene prioritization (Moreau & Tranchevent, 2012) and further select the genes that show significant difference between the two hypertension groups (patient taking 1 vs. multiple anti-hypertensive medications) by two-tailed binomial exact tests (Howell, 2012). Eventu-ally, 349 (110) genes are selected for our cohort with p-value of binomial test less than 0.1 (0.01). Each entry of our genetic matrix specifies how many variants a patient has on that gene.

## 2.5 Hybrid NMF

We propose the hybrid NMF (HNMF) model that integrates both pheno-typic and genetic measurements of patients. The phenotypic measure-ments we consider are continuous values, hence we use Gaussian distri-bution to model the approximation error. The genetic measurements are counts of the genetic variants that happen to a particular gene, thus we use Poisson distribution to model the variant count. A schematic view of our HNMF model is shown in Figure 2.

Table 2: Notations.

| Variable | Description |
|---|---|
| $n$ | number of patients |
| $m_p$ | number of phenotypes |
| $m_g$ | number of genotypes |
| $k$ | number of patient groups |
| $X_p \in R^{n \times m_p}$ | patient by phenotype matrix, continuous value |
| $X_g \in R^{n \times m_g}$ | patient by genotype matrix, count value |
| $F \in R^{k \times n}$ | patient group assignment matrix |
| $G_p \in R^{k \times m_p}$ | phenotype group assignment matrix |
| $G_g \in R^{k \times m_g}$ | genotype group assignment matrix |

Integrating Hypertension Phenotype and Genotype with Hybrid Non-negative Matrix Factorization

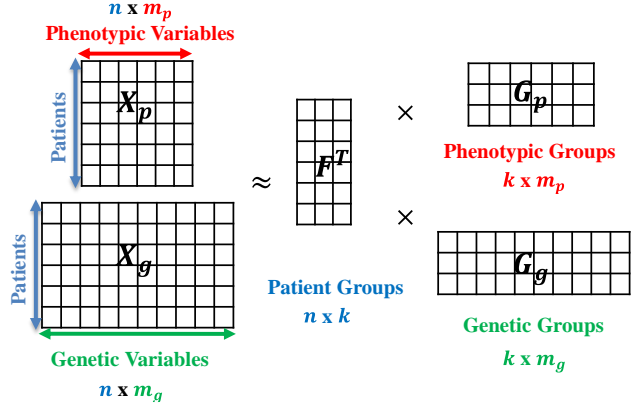

Figure 2: Hybrid Non-negative Matrix Factorization model. In the figure, $X_p$ is the patient-by-phenotype-measurement matrix. $X_g$ is the patient-by-genetic-variant matrix. $F$ is the patient factor matrix specifying patient groups. $G_p$ is the phenotype factor matrix specifying groups of phenotypic measurements. $G_g$ is the genetic factor matrix specifying groups of genetic measurements.

Our goal is to maximize the joint likelihood of the two approximations. Let the variables be defined as in Table 2, we establish the following constrained optimization problem.

$$\max \lambda \log P(X_g|F, G_g) + \log P(X_p|F, G_p) \\ st. F \geq 0, G_p \geq 0, G_g \geq 0 \quad (1)$$

where $\lambda$ indicates the trade-off between the phenotypic approximation and genetic approximation ($\lambda = 1$ for our experiment), and the log likelihood functions are defined as follows.

$$\log P(X_p|F, G_p) = -\frac{1}{2\delta^2} \sum_{ij} \left( X_{p_{ij}} - \sum_u F_{ui} G_{p_{uj}} \right)^2 + C_1 \quad (2)$$

$$\log P(X_g|F, G_g) = \sum_{ij} \left( X_{g_{ij}} \log \left( \sum_u F_{ui} G_{g_{uj}} \right) - \sum_u F_{ui} G_{g_{uj}} \right) + C_2 \quad (3)$$

By minimizing the negative log likelihood, we arrive at the following objective function.

$$\min \mathcal{L}(F, G_p, G_g) = \sum_{ij} \left[ \frac{\lambda}{2} \left( X_{p_{ij}} - \hat{X}_{p_{ij}} \right)^2 + \hat{X}_{g_{ij}} - X_{g_{ij}} \log \left( \hat{X}_{g_{ij}} \right) \right] \\ st. F \geq 0, G_p \geq 0, G_g \geq 0 \quad (4)$$

where $\hat{X}_{p_{ij}} = \sum_u F_{ui} G_{p_{uj}}$ and $\hat{X}_{g_{ij}} = \sum_u F_{ui} G_{g_{uj}}$ Writing $\mathcal{L}$ in the matrix form, we have

$$\mathcal{L}(F, G_p, G_g) = \sum_{ij} \left[ \frac{\lambda}{2} \|X_p - \hat{X}_p\|_F^2 + \hat{X}_g - X_g \log(\hat{X}_g) \right] \quad (5)$$

where $\hat{X}_p = F^T G_p$ and $\hat{X}_g = F^T G_g$. We can use the following alternating projected gradient descent procedure to solve the objective and establish the stopping criteria that the partial gradients should be small enough or all factor matrix updates cannot produce a feasible direction along which the objective function decreases (let $\mathcal{P}_+(\cdot)$ denote the non-negative projector):

$$F^{t+1} = \mathcal{P}_+\left[F^t - \eta_F^t \nabla_F \mathcal{L}(F, G_p^t, G_g^t)|_{F=F^t}\right] \quad (6)$$

$$G_p^{t+1} = \mathcal{P}_+\left[G_p^t - \eta_{G_p}^t \nabla_{G_p} \mathcal{L}(F^{t+1}, G_p, G_g^t)|_{G_p=G_p^t}\right] \quad (7)$$

$$G_g^{t+1} = \mathcal{P}_+\left[G_g^t - \eta_{G_g}^t \nabla_{G_g} \mathcal{L}(F^{t+1}, G_p^{t+1}, G_g)|_{G_g=G_g^t}\right] \quad (8)$$

These equations take turns in optimizing each factor matrix while keeping the other two fixed. We next present the partial gradients with respect to each of the three factor matrices. For phenotype group matrix $G_p$, we have

$$\nabla_{G_p} \mathcal{L}(F, G_p, G_g) = \lambda(FF^T G_p - FX_p) \quad (9)$$

Let $\hat{X}_g = F^T G_g$, and $\tilde{X}_{g_{ij}} = X_{g_{ij}}/\hat{X}_{g_{ij}}$, for genotype group matrix $G_g$, we have

$$\nabla_{G_g} \mathcal{L}(F, G_p, G_g) = F(E_G - \tilde{X}_g) \quad (10)$$

where $E_G \in R^{n \times m_g}$ is an all-one matrix. For the patient group matrix $F$, we have

$$\nabla_F \mathcal{L}(F, G_P, G_g) = \lambda(-G_p X_p^T + G_p G_p^T F) + G_g(E_F - \tilde{X}_g^T) \quad (11)$$

With those gradients, we can adopt an alternating projected gradient descent procedure to solve the hybrid matrix factorization problem. This is an iterative procedure, at each iteration, the algorithm optimizes the objective with one specific group of variables with all other variables fixed. The optimization procedure used at each iteration will be projected gradient descent. In order to determine the step size at each gradient descent step, we use the Armijo rule as a sub-procedure which looks for the largest $\eta$ (step size) that satisfies the following sufficient decrease condition. Let $\Theta$, $\Theta^{new}$ denote the parameters (e.g., $F$, $G_g$ and $G_p$) before and after each iteration respectively, and $\delta \in (0,1)$ be a predefined number. General sufficient decrease condition can be written as

$$\mathcal{L}(\Theta^{new}) - \mathcal{L}(\Theta) \leq \delta tr(\nabla_\Theta \mathcal{L}(\Theta)(\Theta^{new} - \Theta)^T) \quad (12)$$

If $\mathcal{L}$ is a quadratic form of $\Theta$, we have a special fast-to-check sufficient decrease condition as Formula (13) (Lin, 2007). The algorithm for projected gradient descent with Armijo rule can be outlined as Algorithm 1.

$$(1-\delta)tr(\nabla_\Theta \mathcal{L}(\Theta)(\Theta^{new} - \Theta)^T) + \frac{1}{2}tr\left((\Theta^{new} - \Theta)\nabla_\Theta^2 \mathcal{L}(\Theta)(\Theta^{new} - \Theta)^T\right) \leq 0 \quad (13)$$

---

**Algorithm 1 Projected Gradient Descent with Armijo Rule**

1: Initialize $\Theta$. Set $\eta = 1$
2: **for** $i = 1$ to $k$ **do**
3:    **if** $\eta$ satisfies Eq. ( 13 ) (or ( 12 ) if quadratic) **then**
4:       Repeatedly increase $\eta$ as $\eta \leftarrow \eta/\rho$ until either $\rho$ does not satisfy Eq. ( 13 ) (or ( 12 ) if    quadratic) or $\Theta(\eta/\rho) = \Theta$
5:    **else**
6:       Repeatedly decrease $\eta$ as $\eta \leftarrow \rho\eta$ until $\eta$ satisfy Eq. ( 13 ) (or ( 12 ) if quadratic)
7:    **end if**
8:    Set $\Theta^{new} = \max(0, \Theta - \eta\nabla_\Theta \mathcal{L}(\Theta))$
9: **end for**

## 2.6 Feature group discovery using HNMF

In HNMF, the row vectors in the phenotype factor matrix $G_p$ and in the genetic factor matrix $G_g$ specify the grouping of phenotypic measurements and genetic variants respectively. Such groupings can be viewed as mixtures of phenotypic (or genetic) features, as they allow sharing of these features among different groups as specified by its fractional weights across groups. The motivation is to identify paired phenotypic group and genetic group that together characterize pathophysiologic underpinnings. The approximated phenotypic matrix can be viewed as rank-one sum of outer-product of patient group (e.g., $[F^T]_{.5}$, 5th column of the patient group matrix) and phenotypic group (e.g., $[G_p]_{5.}$, 5th row of the phenotypic group matrix). Similar argument holds for genetic group matrix. Thus the patient group (e.g., $[F^T]_{.5}$) bridges the corresponding phenotypic group (e.g., $[G_p]_{5.}$) and genetic group (e.g., $[G_g]_{5.}$). We used the patient group matrix $F^T$ as the instance-feature matrix in Ridge regression, and identify a column with maximum coefficient (e.g., $[F^T]_{.5}$). We selected the corresponding phenotypic and genetic groups (e.g., $[G_p]_{5.}$ and $[G_g]_{5.}$), which are paired through the shared patient group (e.g., $[F^T]_{.5}$) and provide interpretation advantage. Using the trained regression model, we rank the patient groups by their regression coefficients and focus on the top patient groups (and associated phenotypic and genetic groups) that are associated with large effect size.

## 2.7 Evaluating the groups discovered by HNMF

Because there is no innate way (except for simulation) to determine whether the groupings of phenotypic measurements and genetic variants discovered by HNMF are good or poor, we evaluate their utility as features, abstracted from the base data, in a prediction model. We assume that good features will improve prediction and will give us some insights into which phenotypic and genetic patterns are indicative of patient cardiac mechanic abnormality. We use the phenotypic and genetic data for participants from the Hypertension Genetic Epidemiology Network (HyperGEN) study. We take a subset of the African American patients who are hypertensive, and for whom we have both phenotypic and genetic data available at large scale. We predict the numeric values of the cardiac mechanic variables as outcomes (listed in Table 1). For each outcome variable, we randomly split these patients into a 7:3 train and held-out test dataset, and repeat the random initializations of HNMF and other NMF based comparison models 50 times in order to improve the statistical robustness of the results.

To evaluate the effectiveness of HNMF in abstracting raw data into more predictive features, we use the patient factor matrix $F$ to train a Ridge regression model. We chose Ridge regression over alternatives such as support vector regression or random forest regression for its capability to generate deterministic weights for individual features. We match the groups in the phenotypic factor matrix and genetic factor matrix according to their row indices, and link them to the corresponding row in the patient factor matrix $F$. Linear regression then provides a convenient way to directly assess phenotypic and genetic group contribution.

## 3. Results

In this section, we first evaluate the algorithmic performance using a simulated dataset where the actual factor matrices are known. Then, we evaluate the hybrid matrix factorization performance using the HyperGEN dataset.

## 3.1 Simulation

We first analyze simulated data where the underlying factor matrices are known. Specifically, we consider a $20 \times 10$ $X_p$ matrix and a $20 \times 100$ $X_g$ matrix with the true number of factors being 3. That is, they are generated by a $3 \times 20$ $F$ matrix, a $3 \times 10$ $G_p$ matrix, and a $3 \times 100$ $G_g$ matrix. We first sample the $F$, $G_p$, and $G_g$ matrices. We then generate the $X_p$ matrix by adding an error term $\epsilon_p$ on top of $F^T G_p$ where $\epsilon_p$ adopts standard normal distribution. Next we generate the $X_g$ matrix by sampling according to Poisson distribution with the parameter set to $F^T G_g$.

In order to evaluate the similarity between the factorized matrix and its true counterpart, we use the following similarity score:

$$similarity(A, B) = \frac{tr(A^T B)}{\sqrt{tr(A^T A)}\sqrt{tr(B^T B)}} \quad (14)$$

where $tr(\cdot)$ is trace and $tr(A^T B)$ can be considered as matrix inner product. This similarity score is essentially the cosine similarity, which quantifies the closeness between the computed solution and the actual factor matrix and provides a single number between 0 and 1 (Chi & Kolda, 2012). The simulation results are shown in Figure 3(a) where the similarity score is plotted as a function of maximum number of iterations for sub-procedures (optimizing $F$, $G_p$, $G_g$ one at a time while fixing the other two, using the Armijo rule), which represents the closeness to the sub-problem optima. Figure 3 shows that as we have extra sub-procedure iterations, the similarity scores first rise and then plateau quickly. We can also see that the similarity between the true factor matrices and those recovered by HNMF quickly reaches to an accurate level (>0.9). Figure 3(b) shows the convergence speed of the proposed alternating projected gradient descent method with the number of iterations for sub-procedures set to 100. We can see that both loss functions (Frobenius loss for phenotype matrix and KL loss for genotype matrix) quickly decrease within a few iterations. In fact, for our simulation, the stopping criteria is usually met in less than 50 iterations.

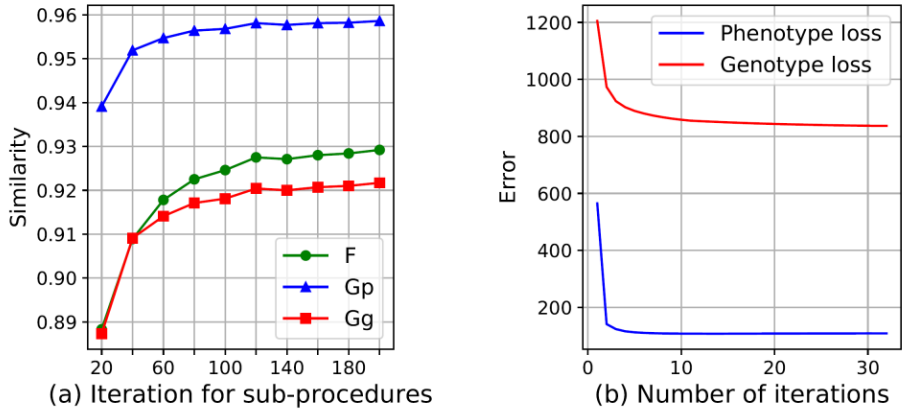

Figure 3: Simulation results on a hybrid matrix factorization problem with rank 3. (a) Similarity score as a function of number of iterations for sub-procedures (b) Decreasing trend of loss functions for phenotype (Frobenius loss) and genotype matrix (KL loss) approximations during HNMF.

### 3.2 Application on cardiac mechanics

We then evaluate HNMF on its effectiveness of abstracting raw data into more predictive features. Using the 2 indices of cardiac mechanics listed in Table 1 as the outcome and the patient factor matrix F as the predictors, we train a Ridge regression model. We evaluate the root-mean-square error (RMSE) of our model on the held-out test set, and compare it against two baselines: (b1)

Using only genetic variants as regression features; (b2) Using only phenotypic measurements as regression features. We also established four comparison models as follows: (c1) Using only the genetic groups as regression features by applying NMF on the genetic variant matrix only; (c2) Using only the phenotypic groups as regression features by applying NMF on the phenotypic measurement matrix only; (c3) Using joint matrix factorization but use KL loss for both matrices; (c4) Using joint NMF but use Frobenius loss for both matrices.

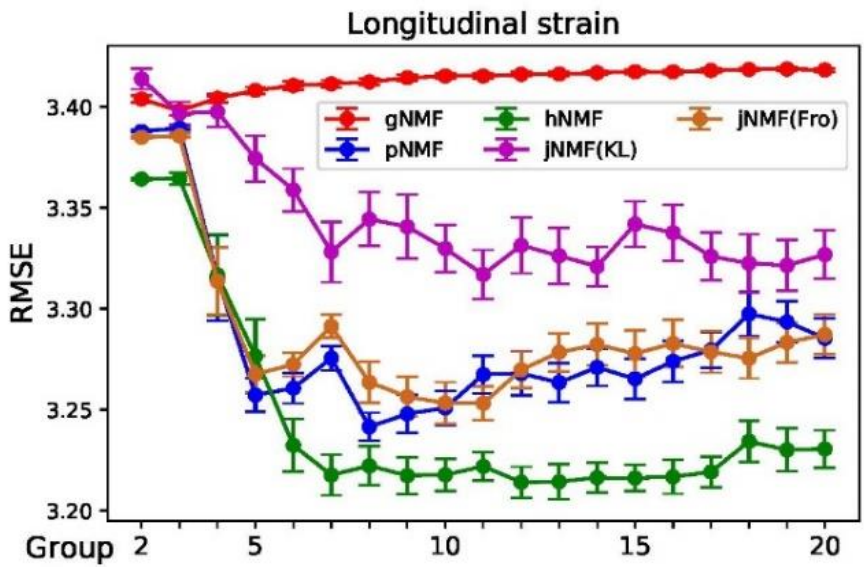

(a)

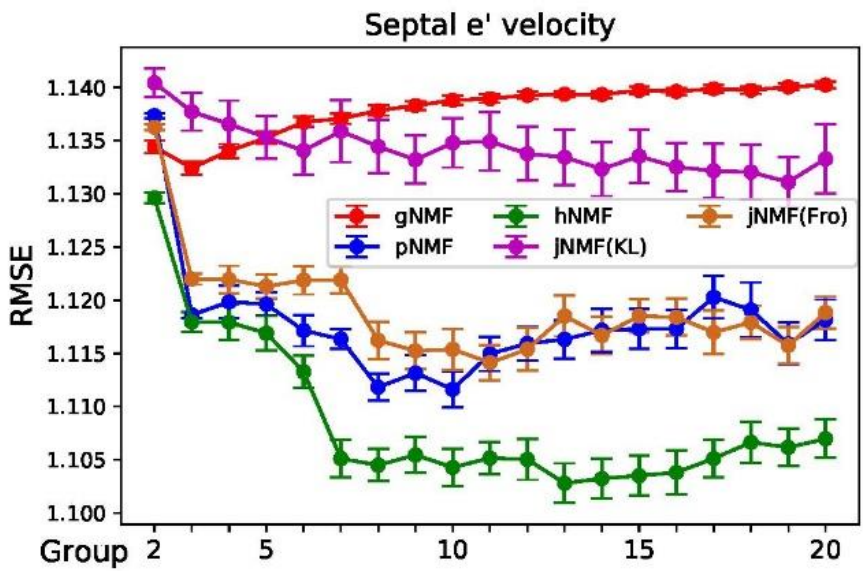

(b)

Figure 4: RMSE with 95% confidence interval for HNMF and comparison methods on the held-out test data. gNMF – using genotype factor matrix as features; pNMF –phenotype factor matrix as features; hNMF –hybrid factor matrix as features; jNMF(KL) – joint matrix factorization using KL loss; jNMF(Fro) – joint matrix factorization using Frobenius loss.

We follow Ho et al. (Ho, Ghosh, & Sun, 2014) on the evaluation procedure in that we vary the group number k from the smallest 2 to where the evaluation metric plateaus and show that across the spectrum HNMF outperforms multiple separate and joint NMF comparison models. The baseline RMSE performances are: 1.25 and 3.88 for geno-baseline on septal e' velocity and longitudinal strain respectively, 1.20 and 3.55 for pheno-baseline respectively. The RMSE performance results of HNMF and comparison models are shown in Figure 4. Comparing all the factorization models and non-factorization models, we can see that using factor matrices as features results in significant improvement (smaller RMSE) over using phenotypic measurements and genetic variants directly as features. Phenotype-only factor matrices often show better regression accuracy than genotype-only factor matrices, likely due to that fact that genetic raw matrix is much sparser than the phenotypic raw matrix. The HNMF factor matrix for regression also significantly outperforms all comparison models including genotype-only or phenotype-only factor matrix for regression, as well as the two joint NMF model results using either KL loss or Frobenius loss for both matrices. This suggests that HNMF can effectively integrate the phenotype and genotype features to predict cardiac mechanics outcomes.

### 3.3 Sensitivity analysis

When performing annotation-based genetic variant filtration, we select the genes that show significant difference in number of LGD variants between the two hypertension groups (patient taking 1 vs. multiple anti-hypertensive medications) by two-tailed binomial exact tests. Using a p-value threshold of being less than 0.01 produces 110 genes for our cohort. This is a relatively stringent threshold and in this section we perform sensitivity analysis by varying the p-value threshold and including 0.05 and 0.1. With these p-value thresholds, we include considerably more genes into consideration: 239 genes for 0.05 as threshold and 349 genes for 0.1 as threshold. The genotype baseline RMSEs are 4.87 (4.63) for longitudinal strain and 1.56 (1.50) for septal e' velocity under p-value threshold 0.1 (0.05). Fig S1 (see Appendix B) shows the results of the sensitivity analysis in comparison with Figure 4. Comparing these figures, it is easy to see that under all p-value thresholds, HNMF consistently outperforms all baselines and NMF comparison models including pheno- and geno- separate NMF models and joint NMF models with KL or Frobenius losses. On the other hand, as one tightens the p-value threshold, the plateau region becomes wider, suggesting that the regression performance is less sensitive as the group number varies in the plateau region. Thus in the following phenotype and genotype group analysis, we chose p-value threshold of 0.01. Another reason is that with a stricter p-value threshold, we are more confident that selected genes are likely implicated in the pathogenesis of abnormal cardiac mechanics. We also note that with large enough patient cohort size, techniques such as cross-validations can be used to accurately determine the optimal group number. The larger the patient cohort size, the more effective cross-validation is, under more relaxed filtering criteria that result in more genes to consider.

### 4. Discussion

Using the method in the feature group discovery section, we identified the top phenotypic and genetic groups that are associated with worse cardiac mechanics. Due to space limitation, we only show the top phenotypic and genetics groups associated with lower values of septal e' velocity and longitudinal strain, as listed in Table 3. The phenotypic groups can help us identify variables that are correlated with abnormal cardiac mechanics. The associated genetic group consists of genes that potentially mediate the corresponding multi-variable phenotypic abnormality. They collectively indicate problematic multi-factor genotype and phenotype interaction and attribute such interaction to a specific patient group (in F), thus can more comprehensively and precisely characterize and stratify these patients in an evidence-driven fashion.

More specifically, the echocardiographic septal e' velocity is one of several variables used during the assessment of diastolic dysfunction. In general lower septal e' values are reflective of a higher degree of diastolic dysfunction, which is associated with the development of heart failure and/or adverse cardiovascular outcomes (Mitter et al., 2017). In septal-e' phenotype group, preserved (higher) left ventricular ejection fraction is often present in patients with diastolic dysfunction, other variables are associated with the development of diastolic dysfunction, including abnormal sodium, calcium, and albumin levels, and abnormal left ventricular wall thickness during diastole. In the septal-e' gene group, TPM2 shows strong susceptibility to variants that lead to cardiomyopathies and IDI2 to chronic kidney disease (comorbidity and risk factor for cardiovascular disease). NPR2 is linked to cardiac conduction. GPRC6A is responsible for calcium sensing that affects L-type calcium channel and is critical to cardiac cell function (Mackenzie et al., 2005). MSMP is involved in resting heart rate modulation. For longitudinal strain, lower values suggest worse longitudinal systolic function of the subendocardium (inner layer of the heart), thus worse cardiac mechanics (Shah et al., 2014). In longitudinal strain phenotype group, besides abnormal sodium, calcium and albumin levels, both higher waist/hip ratio and faster sitting heart rate have a known association with the development of heart failure (Bui, Horwich, & Fonarow, 2011). In the longitudinal strain gene group, COX6B2 is in the cardiac muscle contraction pathway, CLDN5 is expressed in heart muscle, other genes also show strong susceptibility to variants that lead to cardiomyopathies (TPM2), other cardiovascular diseases (BMP4), and obesity as comorbidity (PAX5).

Table 3: Top phenotypic and genetic groups (and their representative components) associated with lower values of septal e' velocity and global longitudinal strain (worse cardiac mechanics). Paired phenotypic group and genetic group are linked by patient group.

| Septal e': Top Phenotype Group | Top Gene Group |
|---|---|
| Sodium | GPRC6A |
| Calcium | MSMP |
| Albumin | NPR2 |
| Left ventricular ejection fraction | IDI2 |
| Relative wall thickness | TPM2 |
| **GLS: Top Phenotype Group** | **Top Gene Group** |
| Sodium | COX6B2 |
| Calcium | PAX5 |
| Albumin | BMP4 |
| waist/hip ratio | TPM2 |
| sitting heart rate | CLDN5 |

To sum, we proposed a novel Hybrid Non-negative Matrix Factorization (HNMF) algorithm that integrates both phenotypic measurements and genetic variants as features in order to subtype patients. HNMF models the approximation error for the phenotypic matrix using Gaussian distribution, and models the variant count for the genetic matrix using Poisson distribution. The objective function is the negative log-likelihood of the data given parameters. We developed an alternating projected gradient descent method to solve the approximation problem. Using the simulated dataset, we demonstrated that HNMF has fast convergence and high accuracy when approximating the true factor matrices. Using the real-world HyperGEN dataset, we demonstrated the effectiveness of HNMF in integrating both the phenotypic and genetic features to derive informative patient subgroupings. We used the patient factor matrix as features to predict the cardiac mechanics outcome variables. We compared HNMF with six different models using phenotype or genotype features directly, using NMF on these features separately, and using joint matrix factorization but with only one type of loss function. HNMF significantly outperforms all comparison models. Analyzing the identified phenotype and genotype groups reveals intuitive phenotype-genotype interactions that characterize cardiac abnormality. For future study, we plan to extend HNMF to consider prior medical knowledge (e.g., known phenotypic and genotypic characteristics associated with heart failure) in guiding the generation of the factor matrices for better patient stratification. We also plan to extend HNMF to a tri-factorization model that allows for different group numbers in patient, genotype and phenotype factor matrices, in order to benefit HNMF with more flexibility to handle heterogeneous and distinct modality of data sources. We plan to model the genetic matrix approximation using zero-inflated Poisson distribution, as genetic matrix is sparse. We also plan to relax LGD criteria to include more genetic variants and obtain Whole Genome Sequencing data to systematically capture potential regulatory variants.

## References

Bui, A. L., Horwich, T. B., & Fonarow, G. C. (2011). Epidemiology and risk profile of heart failure. *Nature Reviews Cardiology, 8*(1), 30-41.

Chi, E. C., & Kolda, T. G. (2012). On tensors, sparsity, and nonnegative factorizations. *SIAM Journal on Matrix Analysis and Applications, 33*(4), 1272-1299.

Collisson, E. A., Sadanandam, A., Olson, P., Gibb, W. J., Truitt, M., Gu, S., . . . Jakkula, L. (2011). Subtypes of pancreatic ductal adenocarcinoma and their differing responses to therapy. *Nature Medicine, 17*(4), 500-503.

DePristo, M. A., Banks, E., Poplin, R., Garimella, K. V., Maguire, J. R., Hartl, C., . . . Daly, M. J. (2011). A framework for variation discovery and genotyping using next-generation DNA sequencing data. *Nature Genetics, 43*(5), 491-+.

Ding, C., He, X., & Simon, H. D. (2005). *On the equivalence of nonnegative matrix factorization and spectral clustering.* Paper presented at the Proceedings of the 2005 SIAM International Conference on Data Mining.

Ding, C., Li, T., Peng, W., & Park, H. (2006). *Orthogonal nonnegative matrix t-factorizations for clustering.* Paper presented at the Proceedings of the 12th ACM SIGKDD international conference on Knowledge discovery and data mining.

Ding, C. H., Li, T., & Jordan, M. I. (2010). Convex and semi-nonnegative matrix factorizations. *IEEE transactions on pattern analysis and machine intelligence, 32*(1), 45-55.

Gunasekar, S., Ho, J. C., Ghosh, J., Kreml, S., Kho, A. N., Denny, J. C., . . . Sun, J. (2016). Phenotyping using Structured Collective Matrix Factorization of Multi--source EHR Data. *arXiv preprint arXiv:1609.04466*.

Guo, T., Yin, R.-X., Pan, L., Yang, S., Miao, L., & Huang, F. (2017). Integrative variants, haplotypes and diplotypes of the CAPN3 and FRMD5 genes and several environmental exposures associate with serum lipid variables. *Scientific Reports, 7*, 45119.

Ho, J. C., Ghosh, J., & Sun, J. (2014). *Marble: high-throughput phenotyping from electronic health records via sparse nonnegative tensor factorization.* Paper presented at the Proceedings of the 20th ACM SIGKDD international conference on Knowledge discovery and data mining.

Hofree, M., Shen, J. P., Carter, H., Gross, A., & Ideker, T. (2013). Network-based stratification of tumor mutations. *Nature methods, 10*(11), 1108-1115.

Katz, D. H., Deo, R. C., Aguilar, F. G., Selvaraj, S., Martinez, E. E., Beussink-Nelson, L., . . . Tiwari, H. (2017). Phenomapping for the identification of hypertensive patients with the myocardial substrate for heart failure with preserved ejection fraction. *Journal of Cardiovascular Translational Research*, 1-10.

Kim, H., Choo, J., Kim, J., Reddy, C. K., & Park, H. (2015). *Simultaneous discovery of common and discriminative topics via joint nonnegative matrix factorization.* Paper presented at the Proceedings of the 21th ACM SIGKDD International Conference on Knowledge Discovery and Data Mining.

Kim, J., & Park, H. (2011). Fast nonnegative matrix factorization: An active-set-like method and comparisons. *SIAM Journal on Scientific Computing, 33*(6), 3261-3281.

Kohane, I. S. (2015). Ten things we have to do to achieve precision medicine. *Science, 349*(6243), 37-38.

Lee, D. D., & Seung, H. S. (1999). Learning the parts of objects by non-negative matrix factorization. *Nature, 401*(6755), 788-791.

Lee, D. D., & Seung, H. S. (2001). *Algorithms for non-negative matrix factorization.* Paper presented at the Advances in neural information processing systems.

Lin, C.-J. (2007). Projected gradient methods for nonnegative matrix factorization. *Neural computation, 19*(10), 2756-2779.

Liu, J., Wang, C., Gao, J., & Han, J. (2013). *Multi-view clustering via joint nonnegative matrix factorization.* Paper presented at the Proceedings of the 2013 SIAM International Conference on Data Mining.

Luo, Y., Szolovits, P., Dighe, A. S., & Baron, J. M. (2016). Using Machine Learning to Predict Laboratory Test Results. *American Journal of Clinical Pathology, 145*(6), 778-788.

Luo, Y., Xin, Y., Joshi, R., Celi, L., & Szolovits, P. (2016). *Predicting ICU Mortality Risk by Grouping Temporal Trends from a Multivariate Panel of Physiologic Measurements.* Paper presented at the Proceedings of the 30th AAAI Conference on Artificial Intelligence.

Mackenzie, P. I., Bock, K. W., Burchell, B., Guillemette, C., Ikushiro, S., Iyanagi, T., . . . Nebert, D. W. (2005). Nomenclature update for the mammalian UDP glycosyltransferase (UGT) gene superfamily. *Pharmacogenetics and Genomics, 15*(10), 677-685.

Mitter, S. S., Shah, S. J., & Thomas, J. D. (2017). A Test in Context E/A and E/e ' to Assess Diastolic Dysfunction and LV Filling Pressure. *Journal of the American College of Cardiology, 69*(11), 1451-1464.

Mor-Avi, V., Lang, R. M., Badano, L. P., Belohlavek, M., Cardim, N. M., Derumeaux, G., . . . Sengupta, P. P. (2011). Current and evolving echocardiographic techniques for the quantitative evaluation of cardiac mechanics: ASE/EAE consensus statement on methodology and indications: endorsed by the Japanese Society of Echocardiography. *Journal of the American Society of Echocardiography, 24*(3), 277-313.

Müller, F.-J., Laurent, L. C., Kostka, D., Ulitsky, I., Williams, R., Lu, C., . . . Schwartz, P. H. (2008). Regulatory networks define phenotypic classes of human stem cell lines. *Nature, 455*(7211), 401-405.

Poulter, N. R., Prabhakaran, D., & Caulfield, M. (2015). Hypertension. *The Lancet, 386*(9995), 801-812.

Selvaraj, S., Martinez, E. E., Aguilar, F. G., Kim, K. Y. A., Peng, J., Sha, J., . . . Shah, S. J. (2016). Association of Central Adiposity With Adverse Cardiac Mechanics Findings From the Hypertension Genetic Epidemiology Network Study. *Circulation-Cardiovascular Imaging, 9*(6).

Shah, S. J., Aistrup, G. L., Gupta, D. K., O'Toole, M. J., Nahhas, A. F., Schuster, D., . . . Wasserstrom, J. A. (2014). Ultrastructural and cellular basis for the development of abnormal myocardial mechanics during the transition from hypertension to heart failure. *American Journal of Physiology-Heart and Circulatory Physiology, 306*(1), H88-H100.

Sra, S., & Dhillon, I. S. (2006). *Generalized nonnegative matrix approximations with Bregman divergences.* Paper presented at the Advances in neural information processing systems.

van Buuren, S., & Groothuis-Oudshoorn, K. (2011). mice: Multivariate Imputation by Chained Equations in R. *Journal of Statistical Software, 45*(3), 1-67.

Wang, H.-Q., Zheng, C.-H., & Zhao, X.-M. (2014). jNMFMA: a joint non-negative matrix factorization meta-analysis of transcriptomics data. *Bioinformatics, 31*(4), 572-580.

Williams, R. R., Rao, D. C., Ellison, R. C., Arnett, D. K., Heiss, G., Oberman, A., . . . Investigators, H. (2000). NHLBI Family Blood Pressure Program: Methodology and recruitment in the HyperGEN network. *Annals of Epidemiology, 10*(6), 389-400.

## Appendix A.

Derivation of the partial gradients

$$\nabla_{G_p}\mathcal{L}(F, G_p, G_g) = \nabla_{G_p}\left[\|X_p - F^T G_p\|_F^2 + C\right] = \nabla_{G_p}\left[tr\left((X_p - F^T G_p)^T(X_p - F^T G_p)\right) + C\right]$$
$$= \nabla_{G_p}[tr(-2X_p^T F^T G_p + G_p^T F F^T G_p) + C] = \lambda(FF^T G_p - FX_p)$$

$$\nabla_{G_g}\mathcal{L}(F, G_p, G_g)\Big|_{ab} = \nabla_{G_g}\left[\sum_{ij}\left(\sum_u F_{ui} G_{g_{uj}} - X_{g_{ij}}\log\left(\sum_u F_{ui} G_{g_{uj}}\right)\right) + C\right]_{ab}$$

Let $\hat{X}_g = F^T G_g$, then

$$\nabla_{G_g}\mathcal{L}(F, G_p, G_g)\Big|_{ab} = \sum_{ij}\left(\frac{\partial \hat{X}_{g_{ij}}}{\partial G_{g_{ab}}} - \frac{X_{g_{ij}}}{\hat{X}_{g_{ij}}}\frac{\partial \hat{X}_{g_{ij}}}{\partial G_{g_{ab}}}\right)$$

We know $\dfrac{\partial \hat{X}_{g_{ij}}}{\partial G_{g_{ab}}} = F_{ai}$

When $j = b$, let $\tilde{X}_{g_{ij}} = X_{g_{ij}}/\hat{X}_{g_{ij}}$, for genotype group matrix $G_g$, we have

$$\nabla_{G_g}\mathcal{L}(F, G_p, G_g)\Big|_{ab} = \sum_i (F_{ai} - \tilde{X}_{g_{ib}} F_{ai}) = (F\mathbf{1})_a - (F\tilde{X}_g)_{ab}$$

So $\nabla_{G_g}\mathcal{L}(F, G_p, G_g) = F(E_G - \tilde{X}_g)$, where $E_G \in R^{n \times m_g}$ is an all-one matrix.

Now let's compute the gradient with respect to $F$, which includes two parts

$$\nabla F\|X_p - F^T G_p\|_F^2 = 2(-G_p X_p^T + G_p G_p^T F)$$

$$\sum_{ij}\left(\frac{\partial \hat{X}_{g_{ij}}}{\partial F_{ab}} - \frac{X_{g_{ij}}}{\hat{X}_{g_{ij}}}\frac{\partial \hat{X}_{g_{ij}}}{\partial F_{ab}}\right) = (G_g \mathbf{1})_a - (G_g \tilde{X}_g^T)_{ab}$$

Therefore, we have

$$\nabla_F \mathcal{L}(F, G_P, G_g) = \lambda(-G_p X_p^T + G_p G_p^T F) + G_g\left(E_F - \tilde{X}_g^T\right)$$

where $E_F \in R^{m_g \times n}$ is an all-one matrix.

Integrating Hypertension Phenotype and Genotype with Hybrid Non-negative Matrix Factorization

**Appendix B.**

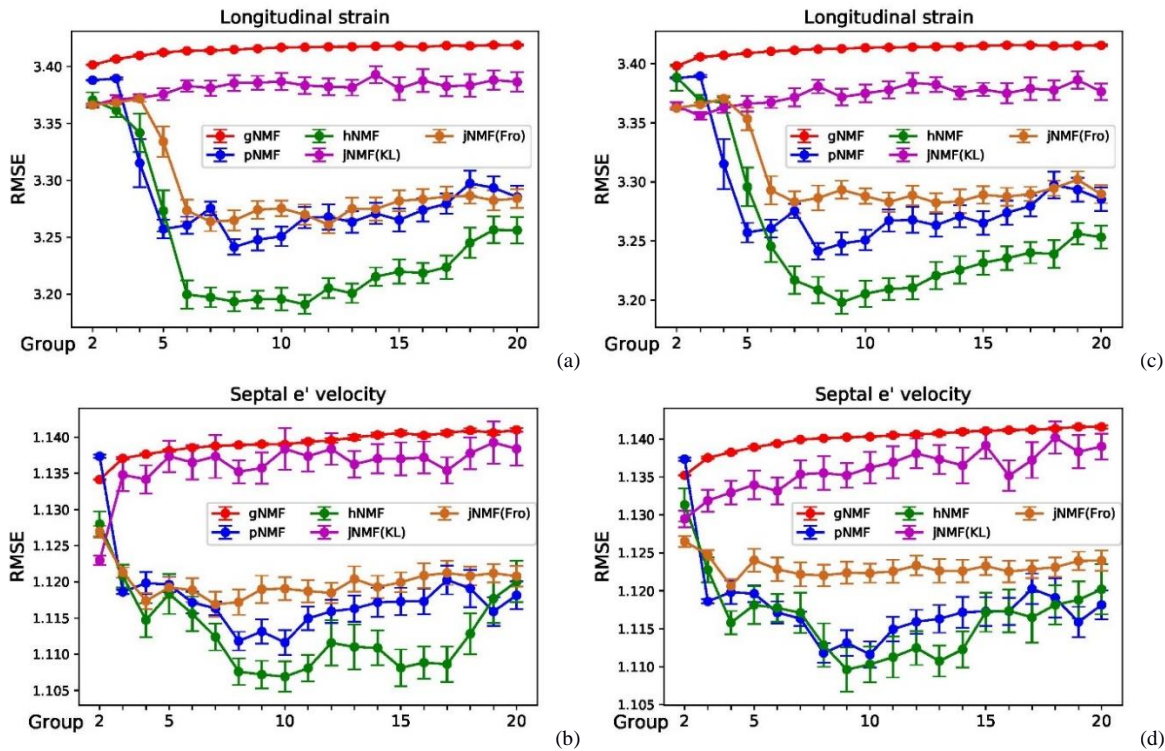

Fig. S1. Sensitivity analysis by varying the gene filtering p-value threshold. (a) and (b) correspond to using $p<0.05$ as threshold (239 genes); (c) and (d) correspond to using $p<0.1$ as threshold (349 genes). The genotype baseline (no NMF) RMSEs are 4.87 (4.63) for longitudinal strain and 1.56 (1.50) for septal e' velocity under p-value thresh-old 0.1 (0.05). Abbreviation: root-mean-square error (RMSE)